# Development of a gaseous proton-recoil detector for neutron flux measurements between 0.2 and 2 MeV neutron energy

Paola Marini[a], Ludovic Mathieu[a], Mourad Aïche[a], Serge Czajkowski[a], Beatriz Jurado[a], Igor Tsekhanovich[a]

[a]CENBG, CNRS/IN2P3-Université de Bordeaux 19, Chemin du Solarium, 33175 Gradignan, France

**Abstract**

Absolute measurements of neutron fluence are an essential prerequisite of neutron-induced cross section measurements, neutron beam lines characterization and dosimetric investigations. The H(n,p) elastic scattering cross section is a very well known standard used to perform precise neutron flux measurements in high precision measurements. The use of this technique, with proton recoil detectors, is not straightforward below incident neutron energy of 1 MeV, due to a high background in the detected proton spectrum. Experiments have been carried out at the AIFIRA facility to investigate such background and to determine its origin and components. Based on these investigations, a gaseous proton-recoil detector has been designed with a reduced low energy background. A first test of this detector has been carried out at the AIFIRA facility, and first results will be presented.

## 1. Introduction

Five out of the six concepts of new generation nuclear reactor studied in the framework of the Generation IV international forum are based on fast neutron spectra. These concepts rely on simulations based on high quality nuclear data (Aliberti et al., 2006). It is expected that the evolution of the GenIV system analysis, and the focusing on a few preferred concepts, will motivate designers to define the relevant target accuracy. Despite the experimental efforts, large discrepancies and uncertainties still exist in the actinide region (Chadwick et al., 2011). Fission cross sections are of primary importance to determine reactor behavior and incineration capabilities. As fast neutron spectra are peaked between 100 keV and 2 MeV, new cross section measurements have to be performed in this energy range (NEA, 2016).

For neutron-induced fission cross section measurements, the accurate knowledge of the neutron flux impinging on the studied sample is mandatory. This is usually performed exploiting a well-known standard reaction such as $^{235}$U(n,f), $^{238}$U(n,f) or $^{237}$Np(n,f). However, this introduces a strong correlation between independent measurements based on the same standard. In addition, the accuracy of these standards is limited (0.5% at best, and up to 10%). An independent and well known standard is the H(n,p) elastic scattering reaction. Indeed, the detection principle is completely different from the fission detection, nullifying correlations between the measured cross section and the standard one. Moreover, its accuracy is about 0.2% in the MeV energy range, compared with typically few percents for the fission cross section standards mentioned above. In this method, the neutron flux is converted into a proton-recoil flux, by mean of a H-rich foil. The charged protons can be easily detected, for instance in a Silicon detector. Different concepts of proton-recoil detectors are available, depending on the type of measurements, the facility environment, the neutron energy range and the target accuracy of the measurement (Hassard et al., 1998, Asai et al., 2006, Babut and Gressier, 2007, Beyer et al., 2007, Donzella et al., 2010, Watanabe et al., 2011, Kovash et al., 2011, Kessedjian et al., 2012, Taforeau et al., 2014).

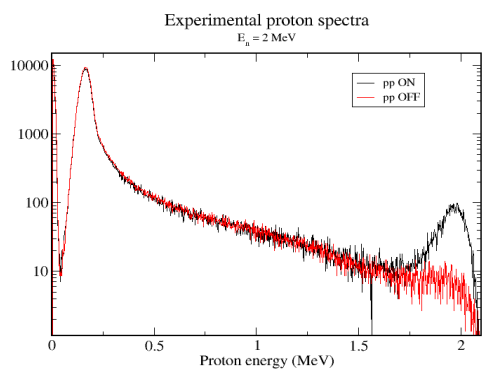

Fig. 1. Proton spectrum in the Si detector (standard and background measurements) for 2MeV incident neutron energy. The proton peak is visible on the right, as well as the intense background at low energy.

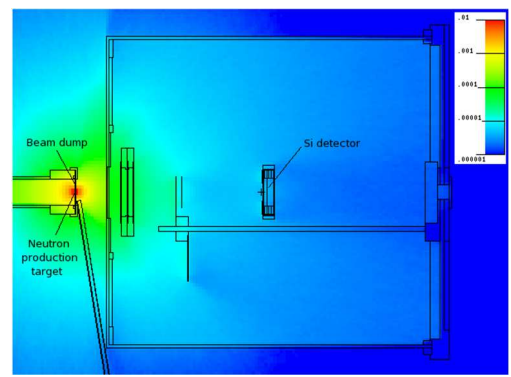

Fig. 2. MCNP simulation of the electron flux, expressed in part/cm²/s for one source particle, in the experimental setup. These electrons come from interactions of 1MeV γ-rays (emitted by the neutron production source) with surrounding materials.

## 2. Proton-recoil detection with Si detector below 1 MeV

A cross section measurement was carried out by the ACEN group of the CENBG institute between 1 and 2 MeV neutron energy, produced via T(p,n) nuclear reaction at 4MV Van De Graaff facility at Bruyères-Le-Châtel, CEA, France. The proton-recoil detector consists of a collimated 4μm-thick polypropylene foil (PP foil) and a collimated 50μm-thick Si detector placed 8 centimeters far away. The protons detected in the silicon are emitted close to 0° with respect to the incoming neutron direction, and have therefore a kinetic energy close to the incident neutron energy. It produces a peak in the proton-recoil spectrum, with a good signal-to-noise ratio. Therefore, the number of counts in the proton peak is used to infer the incident neutron flux. A detailed description of the method and of the detector itself can be found in Marini et al. (2016, 2017). Previous studies at higher energies (using a telescope instead of a single Si detector) showed that a precision of the order of 1% could be achieved in the proton peak integral (Kessedjian et al., 2012). Around 1 or 2 MeV, the accuracy is only about 3% due to a high contribution in the low energy part of the proton spectrum, as can be seen in Fig. 1. Below 1 MeV, it becomes impossible to infer the neutron flux with acceptable uncertainties. Dedicated experiments were performed on the AIFIRA facility to investigate the origin of such background and to find solutions to reduce it. Simulations have also been carried out with the MCNP code, in order to improve our understanding of the observed phenomena.

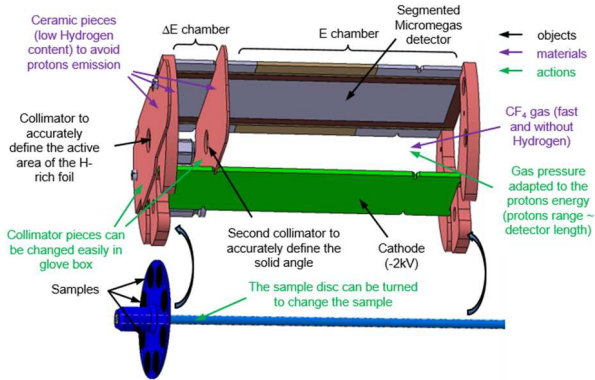

Fig. 3. Scheme of the GPRD. Objects, materials and actions are detailed. The sample disk is shown separately for clarity purpose.

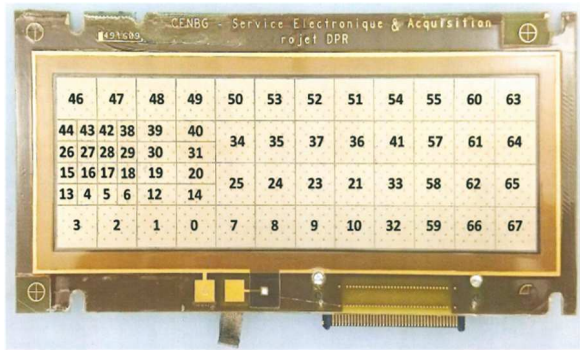

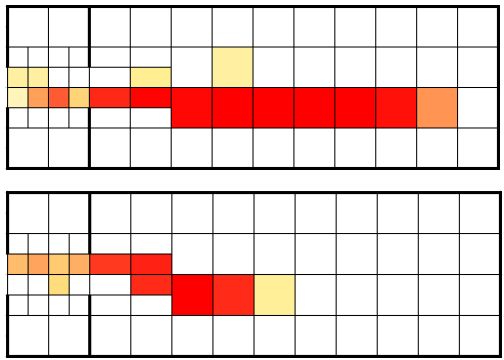

Fig. 4. (up) Picture of the segmented Micromegas detector; (middle) scheme of a proton track for a 1 MeV incident neutron energy; (down) scheme of a proton track for a 300 keV incident neutron energy

The results of the MCNP simulations show that low energy γ-rays emitted at the neutron production source are responsible for the observed background. Indeed, they generate a large amount of electrons via Compton scattering or photoelectric effect when interacting with surrounding materials in the setup (especially the copper beam dump). Fig. 2 shows the electron flux generated by a 1MeV γ-rays source located on the beam dump. These γ-rays are themselves generated by neutron radiative capture reactions, especially in the beam dump where the neutron flux is the highest. Thus, the high γ-rays/electron flux is present as soon as neutrons are produced. Electrons deposit much less energy in matter than protons due to their lower stopping power. Nevertheless, an electron of few 100 keV deposits energy in the whole thickness of the detector (here 50 µm), whereas the ranges of 700 keV and 200 keV protons-recoil in silicon are around 10 µm and 2 µm respectively. In addition, simulations show that many electrons reach the Si detector with a high incident angles, further increasing the effective thickness and thus the energy deposition. Finally, the high electron flux induces pile-up events in the Si detector, contributing to the exponential-shaped high-energy part of the background.

## 3. Development of the Gaseous Proton-Recoil Detector (GPRD)

To perform neutron flux measurement at energies below 1 MeV, the ACEN group has developed a new concept of proton-recoil detector with a reduced sensitivity to γ-ray/electron flux. To adapt the thickness of the detector to the proton energy, the choice was made to build a gaseous detector. The pressure can be adjusted for the total proton range in the gas to fit the detector geometry. The detector has the capability to discriminate direct neutrons from scattered neutrons, as well as to reject contaminant events (protons emitted by other sources than PP foil, cosmic rays…), thanks to a proton track analysis. To perform accurate neutron flux measurements, the detector requires a well-defined efficiency. A scheme of the detector is presented in Fig. 3.

The detector is constituted by a H-rich polypropylene or mylar foil and a segmented ΔE-E ionization chamber (4x4x12cm$^3$). The chamber is filled with $CF_4$ gas with a pressure ranging from few tens of mbar to slightly above 1 bar. The gas has been chosen free of hydrogen, to prevent additional recoil protons emission from the gas itself. Structures and collimators of the GPRD are made of Macor ceramic for the same reason. A special care was taken to build the detector with a reduced amount of material, in order to minimize the neutron scattering which seriously complicates the cross section measurement. The coincidence between the ΔE and E sides of the ionization chamber enables a first discrimination between direct and scattered neutrons to be made. Indeed the former impinge on the H-rich sample with small angles and maximum energy, whereas the latter lose energy in previous collisions and reach the H-rich sample with larger angles. Different H-rich samples can be mounted on a Macor rotating disk at the entrance of the detector. The rotation of the disk step-by-step allows one to change and position the samples without opening the chamber. An empty slot is available for background measurements. The possibility to select the sample thickness adapted to the neutron energy is crucial to perform measurements on a wide energy range (from few 100 keV to few MeV).

The detection plane consists in a Micromegas detector segmented in 64 pads as shown in the upper part of Fig. 4. The Micromegas is able to multiply the primary electrons generated by the recoiling proton inside the detection gas by a factor up to $10^6$. Indeed, the number of primary electrons can be as low as 1000 electrons per chamber side for a 200 keV proton. The segmentation of the detection plane enables both the reduction of the electronic noise associated to the pad capacitance, and a raw track analysis to be performed, in order to reject accidental coincidences with abnormal trajectories.

The detector prototype was delivered at mid 2016 and a first experiment was carried out at the AIFIRA facility to evaluate its response and performances. Measurements were performed with quasi-monoenergetic neutron beam, with energies from 0.3 to 1 MeV, and with a 3α source mounted on the sample disk. The gas pressure was varied from 30 to 100 mbar, depending on the neutron energy, and the applied high voltage was adapted consequently. On the first hand, this experiment validated several expected characteristics.

<u>Proton track analysis:</u> as shown in the middle part of Fig.4, proton tracks were detected with sufficient precision to perform abnormal tracks rejection. In particular, proton emitted by scattered neutrons (which have lower energies) have a shorter range and can be rejected. Cosmic rays can also be seen, and rejected via track analysis.

Low sensitivity to e⁻ or γ-rays: tests carried out without the PP film show no response of the pads. Indeed each pad is not sensitive enough to detect the low energy deposited by electrons.

Low detection energy limit: neutrons with energy down to 300 keV have been produced by the accelerator. At such low energy, as the gas pressure could not be reduced below 30 mbar because of discharge issues in the detection gas, the proton ranges are only of few centimeters. As shown in the lower part of Fig.4, proton tracks are still visible under these conditions.

These different features validate the new principle of this detector. However, under the present experimental conditions, it was not possible to measure the proton detection efficiency of the detector, and in particular that every acceptable recoiling proton is actually detected. It would have required a quantitative experiment, with the addition of a reference fissile target.

On the other hand, this experiment shed light on several issues, which prevent considering the GPRD to be operational in its current status.

Static discharges: at few tens of mbar of gas pressure, the high voltage breakdown is strongly reduced due to the Paschen law (Škoro, 2012, Marić et al., 2014). This effect was under-estimated for the design of the detector, and discharges occurred at certain regimes (pressure/voltage). Special care will be taken to modify the detector design according to this constraint.

Non-homogeneous electric field: The detected signal amplitude of the side pads was quite lower when compared to the central pads. Moreover, the positioning of the collimators (see Fig. 3) also affected the signal amplitude of the neighboring pads. This is due to field lines not strictly perpendicular to the Micromegas and the cathode. Some of the primary electrons then drift along the field lines out of the Micromegas detection plane. To solve this issue, a field cage will be designed for the GPRD upgrade.

Signal weakening: Under neutron irradiation, the signal amplitude of many pads quickly decreases. As can be seen on Fig. 5, the signal is nearly lost after few minutes of irradiation. As said in section 2, a large number of low energy electrons are generated with the neutron production source. These electrons may be deposited on the surfaces of the collimators or samples disk. This induces a strong local electric field distortion, and greatly reduces the primary electrons collection by the pads or affect their spatial distribution. Indeed, by electrically grounding the entrance collimator, the signal loss was less severe (see Fig. 5) as part of the background electrons deposited on the collimator surface are collected and removed. The addition of a field cage will solve this issue, as it will help to collect the parasitic charges and reduce the electric field lines distortion.

## 4. Conclusion

The use of the H(n,p) standard reaction to accurately determine the neutron flux below 1 MeV is challenging because of the high background contribution, mostly due to γ-rays/electrons produced in the neutron production target. To overcome this issue, a new double chamber gaseous proton-recoil detector was designed and tested at the CENBG. It has a very low sensitivity to γ-rays/electrons, which allows to detect recoiling protons generated by incident neutron energies as low as 300 keV. The ΔE-E coincidences, coupled to a proton track analysis, enables the disentanglement between direct neutrons impinging on the PP foil, scattered neutrons impinging on the PP foil, and direct neutrons impinging on other H-rich surrounding materials.

However, the electric field is slightly distorted by surrounding materials, and heavily distorted by the space charge accumulation on ceramic pieces of the detector. New studies based on electrostatic simulation code to add a field cage are undertaken in order to overcome both issues. A new prototype is currently designed and will be tested in the near future.

**Acknowledgement**

This work was supported by the European Commission within the Seventh Framework Programme through Fission-2013-CHANDA (project no.605203).

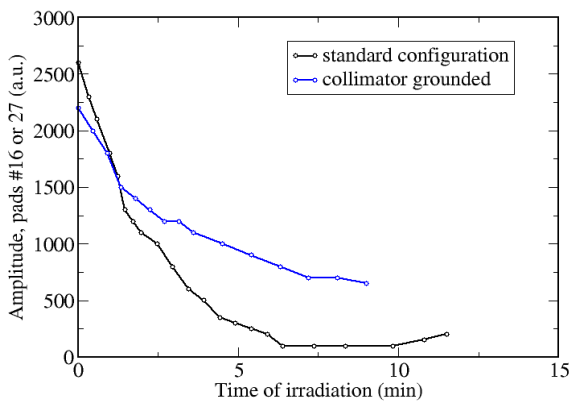

Fig. 5. Signal amplitude on pads in the ΔE region as a function of the time of irradiation